\documentclass[prl,showpacs,bibnotes]{revtex4}
\usepackage{graphicx}
\begin{document}
\draft
\def\ds{\displaystyle}
\title{Free electron in a laser field: The nonrelativistic solution}
\author{C. Yuce }
\address{Department of Physics, Anadolu University,
 Eskisehir, Turkey}
\email{cyuce@anadolu.edu.tr}
\date{\today}
\begin{abstract}
Schrodinger equation for a charged particle interacting with the
plane wave electromagnetic field is solved exactly. The exact
analytic solution and the perturbative solution  up to second
order are compared.
\end{abstract}
\maketitle The exact solution of Dirac's equation for an electron
in an external plane wave electromagnetic field was first obtained
by Volkov \cite{volkov} in 1935 and it's Green function was
derived by Schwinger \cite{schwinger}. The state of the charged
particle, known as Volkov state, has been used extensively to
explore numerous quantum phenomena such as Compton scattering,
photo-ionization, bremsstrahlung processes and Kapitza-Dirac
problem (scattering of an electron by a standing light wave), etc.
However, an exact analytic solution of Schrodinger equation for a
charged particle interacting with a plane wave electromagnetic
field has never been obtained. The subject of the interaction of
the charged particle with radiation is investigated using
perturbation theory in almost any quantum mechanics textbooks.
However, the behavior of a charged particle in the strong
radiation fields by high-power laser can not be investigated by
perturbation theory. To deal with this problem, new approximation
methods have been developed
\cite{ek1,ek11,ros,reiss,gelt,ek2,ekle}. For example, in dipole
approximation, electromagnetic field is assumed to be purely
time-dependent for the study of ionization of atoms by intense
laser pulses. In this approach, the magnetic field is neglected.
As opposes to dipole approximation, the effect of magnetic field
can be taken into account by expanding the potential to first order in the space coordinate.\\
In this paper, we will solve exactly the Schrodinger equation for
an electron in a linearly polarized monochromatic plane wave
propagating in the z-direction. The vector potential is given by
$\ds{\textbf{A}=(A_0 ~e^{ik(z-ct)},~0,~0)}$. The laser pulse with
the elliptic polarization will be discussed later.
\begin{equation}\label{1}
i\hbar \frac{\partial \Psi}{\partial t}=\frac{\textbf{p}^2}{2m}
\Psi-\frac{e}{mc} \textbf{A}. \textbf{p}  \Psi +\frac{e^2}{2mc^2}
\textbf{A}^2 \Psi ~,
\end{equation}
where $\ds{\nabla. \textbf{A}=0}$. To solve this equation, let us
transform z- coordinate as $s=z-ct$. Then the time derivative
operator transforms as $\ds{\partial/\partial t \rightarrow
\partial/\partial t-c\partial/\partial s}$. The equation (\ref{1}) becomes
\begin{equation}\label{21}
i\frac{2m}{\hbar }(\frac{\partial \Psi}{\partial
t}-c\frac{\partial \Psi}{\partial s})=- \nabla^2 \Psi+2i\alpha~
e^{iks} ~\frac{\partial \Psi}{\partial x} +\alpha^2 ~e^{2iks} ~
\Psi ~,
\end{equation}
where $\alpha$ is defined as $\ds{\alpha=\frac{e A_0}{\hbar c}}$
and $\ds{\nabla^2= {\partial_x^2}+{\partial_y^2}+{\partial_s^2}}$.
We introduce the following ansatz; i.e. we look for the solution
of (\ref{21}) in the form
\begin{equation}\label{3}
 \Psi (x,y,s,t)=~\exp{\left(-i\frac{\epsilon}{\hbar} t-i{ {k^{\prime}_x} } x-i {{ {k^{\prime}_y} }} y \right)}
\Phi(s)~,
\end{equation}
where $\epsilon$ and $ {k^{\prime}_x} ,  {k^{\prime}_y} $ are all
constants. If we substitute (\ref{3}) into the equation
(\ref{21}), we obtain
\begin{equation}\label{4}
- \frac{d^2 \Phi}{d s^2}+ i\frac{2mc}{\hbar }\frac{d \Phi}{d
s}+\left(( {k^{\prime}_y} ^2+ {k^{\prime}_x}
^2-\frac{2m}{\hbar^2}\epsilon)+ 2\alpha  {k^{\prime}_x}  ~
e^{iks}+\alpha^2 ~e^{2iks} \right) ~ \Phi=0 ~.
\end{equation}
Let us make a complex transformation on the coordinate as
$\ds{u=\exp(iks)}$. This is reasonable since the Hamiltonian is
not Hermitian. Using the transformation for the derivative
operator $\ds{\frac{d}{d s}=iku\frac{d}{d u}}$, the equation
(\ref{4}) becomes
\begin{equation}\label{6}
u^2 \frac{d^2 \Phi}{du^2}+(1- 2\sigma)u \frac{d
\Phi}{du}+\left(\delta+ \frac{2\alpha  {k^{\prime}_x} }{k^2}~
u+\frac{\alpha^2}{ k^2} ~u^2  \right) ~ \Phi=0 ~,
\end{equation}
where $\ds{\sigma=\frac{mc}{\hbar k}}$, $\ds{\delta=\frac{
{k^{\prime}_y} ^2+ {k^{\prime}_x} ^2-\frac{2m}{\hbar^2}
\epsilon}{k^2}}$. To solve  (\ref{6}), the last transformation is
introduced as
\begin{equation}\label{3333}
 \Phi (u)=\exp{(-i\frac{\alpha u}{ k} )} ~u^{\gamma} ~L(u)~,
\end{equation}
where $\ds{\gamma}$ is a constant to be determined later and
$L(u)$ is a function depending on $u$. If $\gamma$ satisfies a
polynomial equation $\ds{\gamma^2-2\sigma \gamma+\delta=0}$, then
$\ds{L (u)}$ satisfies
\begin{equation}\label{7}
u^2 \frac{d^2 L}{d
u^2}+\left((2\gamma+1-2\sigma)u-2i\frac{\alpha}{k}
u^2\right)\frac{d L}{d u}+\left(2\frac{\alpha  {k^{\prime}_x} }{
k^2}-i\frac{\alpha}{k}(2\gamma+1-2\sigma) \right)u ~ L=0 ~.
\end{equation}
The solution is given in terms of the associated Laguerre
functions $\ds{ L_{\sigma-\gamma-i \frac{ {k^{\prime}_x}
}{k}-\frac{1}{2}}^{2\gamma-2\sigma} \left(\frac{2i\alpha u}{k}
\right)}$. Note that a generalization of the associated Laguerre
polynomial to not necessarily an integer is called the associated
Laguerre function. Transforming backwards yields the exact
solution
\begin{equation}\label{cozum}
\Psi =\exp{\left(-i\frac{\epsilon}{\hbar} t-i {{ {k^{\prime}_x} }}
x-i {{ {k^{\prime}_y} }} y -i\frac{\alpha}{k}e^{ik(z-ct)}+i \gamma
k(z-ct)\right)}~L_{\sigma-\gamma-i \frac{ {k^{\prime}_x}
}{k}-\frac{1}{2}}^{2\gamma-2\sigma} \left(\frac{2i\alpha
e^{ik(z-ct)}}{k} \right)~.
\end{equation}
To determine $\ds{\epsilon}$ and $\gamma$, let us set the charge
to zero $\ds{(\alpha=0)}$. In this case, the free particle
solution should be recovered. Note that Laguerre function is a
number at zero and the constants can be omitted since the free
particle solution isn't square integrable.  \\
To obtain the free particle solution, let us choose $\ds{
\gamma=-{ {k^{\prime}_z} }/k }$. Then, by solving the polynomial
equation for $\gamma$, we get $\ds{-\delta=\frac{{ {k^{\prime}_z}
}}{k}(\frac{{ {k^{\prime}_z} }}{k}+2\sigma)}$. Using the
definition of $\delta$, we obtain $\ds{\epsilon-\hbar
{k^{\prime}_z} c=\frac{\hbar^2}{2m} ( {k^{\prime}_x} ^2+
{k^{\prime}_y} ^2+ {k^{\prime}_z} ^2)}$. Defining a new constant
as $\ds{E=\epsilon+\gamma \hbar k c=\frac{\hbar^2}{2m}
{{\vec{{k}^{\prime}}}}^2}$, the free particle solution is
recovered. Physically, $E$ is the energy for the free particle.
Substituting $\ds{\gamma, \delta, \epsilon, E}$ in (\ref{cozum}),
we get
\begin{equation}\label{cozum20}
\Psi(x,y,z,t) =~\exp{\left(-i\frac{E}{\hbar} t-i
\vec{{{k^{\prime}}}}.
\vec{r}-i\frac{\alpha}{k}e^{ik(z-ct)}\right)}~L_{\sigma-\frac{1}{2}+\frac{
{k^{\prime}_z}-i{k^{\prime}_x} }{k}}^{-2\frac{ {k^{\prime}_z}
}{k}-2\sigma} \left(\frac{2i\alpha e^{ik(z-ct)}}{k} \right)~.
\end{equation}
This is the exact solution of Schrodinger equation with the plane
wave electromagnetic field polarized in the x-direction. If we set
$\alpha$ to zero, then it is reduced to the free particle
solution. The first two terms in the exponential is the exact
solution in the absence of the external field, the other two terms
is due to the interaction with the field.\\
The solution has been obtained for the linearly polarized
electromagnetic field. If the field is elliptically polarized,
then the term $\ds{ {k^{\prime}_x}}$ in Laguerre function should
be replaced by the linear combination of $\ds{ {k^{\prime}_x},
{k^{\prime}_y}}$. \\
Finally, a few remarks about the solution (\ref{cozum20}) is in
order. Some approximations can be made for the associated Laguerre
function $\ds{ L_{\sigma-\frac{1}{2}+\frac{
{k^{\prime}_z}-i{k^{\prime}_x} }{k}}^{-2\frac{ {k^{\prime}_z}
}{k}-2\sigma} }$. The order of $\sigma$ is only comparable to the
factor $1/2$ when $k \approx 10^{11}~ m^{-1}$ (x-rays). Otherwise,
one of them can be neglected. Furthermore, if $k^{\prime}$  is
small compared to $k$, then this term can be neglected, too.\\
As a special case, perturbative and our non-perturbative solutions
should yield the same results. To check the validity of our
result, let us compare the exact solution (\ref{cozum20}) with the
the one obtained by perturbation theory. Using the series
expansion of associated Laguerre function
$\ds{L_n^m(x)=\frac{\Gamma(m+n+1)}{\Gamma(m+1) \Gamma(n+1)}
\left(1-\frac{nx}{m+1}-\frac{(1-n) n
x^2}{2(m+1)(m+2)}-...\right)}$, where $\Gamma$ is the gamma
function, the wave function (\ref{cozum20}) can be expanded in
series up to order $\ds{\alpha^3}$ as
\begin{equation}\label{expansion}
\Psi=e^{-i(k^{\prime}_x x+k^{\prime}_y y)}\left(e^{-ik^{\prime}_z
z-\frac{iEt}{\hbar}}+ \alpha~ c_1
e^{-i(k^{\prime}_z-k)z-\frac{i(E+\hbar kc)t}{\hbar}}+\alpha^2~
c_2e^{-i(k^{\prime}_z-2k)z-\frac{i(E+2\hbar kc)t}{\hbar}} +...
\right)
\end{equation}
where $\ds{c_1=\frac{2k^{\prime}_x }{k (2k^{\prime}_z+k
(2\sigma-1))} }$,~ $\ds{c_2=\frac{(2k
k^{\prime}_z+4{k^{\prime}_x}^2+k^2 (2\sigma-1))}{4k^2
(k(\sigma-1)+k^{\prime}_z)(2k^{\prime}_z+k(2\sigma-1))}}$~. The
constant in front of this expansion can be dropped
from the solution since it is not physical.\\
Having obtained the series expansion of the exact solution (9),
let us review the perturbative expansion for the problem. Assuming
$\alpha$ is small, perturbation theory states that $\Psi$ can be
approximated as
\begin{equation}\label{perturbexpansion}
\Psi=\psi_0+\alpha ~b_1  \psi_1+\alpha^2 ~b_2 \psi_2+O(\alpha^3)~,
\end{equation}
where $\ds{\psi_0}$ is the free particle solution. Substituting
the equation (\ref{perturbexpansion}) into the corresponding
Schrodinger equation (\ref{1}) yields~ $\ds{ \alpha~ b_1~
(\frac{\textbf{p}^2}{2m}\psi_1-i\hbar \dot{\psi}_1)=\frac{e}{mc}
\textbf{A} \textbf{p}~ \psi_0}$ for the first order of $\alpha$.
Furthermore,  for the second order $\ds{ \alpha^2}$, we are left
with the following equation ~$\ds{\alpha^2  b_2~
(\frac{\textbf{p}^2}{2m}\psi_2-i\hbar \dot{\psi}_2)=b_1 \alpha
\frac{e}{mc} \textbf{A} \textbf{p} ~\psi_1-\frac{e^2}{2mc^2}
{\textbf{A}}^2 \psi_0}$. By solving these two equations, we see
that $\ds{b_1=c_1}$ and $\ds{b_2=c_2}$. Then, it is concluded that
our non-perturbative exact
solution is in agrement with the solution obtained by using the perturbation theory. \\
Perturbation theory fails when applying to the system with strong
electromagnetic fields. In contrast, the solution (\ref{cozum20})
is non-perturbative exact solution and it can be applied without
restricting ourselves to the weak laser field.\\
The non-perturbative exact solution obtained here can be used in
many branches of physics. For example, the recent development of
high power laser delivering pulses of intensity up to ${\ds
10^{22} }$ W/$cm^2$ enables us to investigate deeply the
properties of atoms, molecules, plasmas and condensed matter
interacting with super intense laser pulses. With rising laser
intensity, non-perturbative treatments are required to study the
electronic motion of free and bound particles in intense laser
fields. Analytical studies of strong laser atom interaction have
seen much progress over  the years
\cite{kfl,hakem1,hakem2,duz1,duz2}. A successful theory so called
strong field approximation (SFA) was developed by Keldysh, Faisal
and Reiss \cite{kfl}. In SFA, the final state of the ionized
electron is represented by Volkov solution which describes a free
electron interacting with laser field. This is because laser field
dominates Coulomb interaction. With this approximation, the
physical phenomenons such as above-threshold ionization and high
harmonic generation have been explained successfully. Exact
solution for Coulomb field is well-known and given in terms of
associated Laguerre polynomials. The exact solution for an
electron interacting with radiation is given in the equation
(\ref{cozum20}). Thus, for the problem of atom-laser interaction,
either field dominates the other can be treated perturbatively.\\
The dipole approximation is often be employed in atomic physics.
If the Bohr radius $\ds{a_0}$ is very smaller than the wavelength
of the radiation $\ds{a_0<<\lambda}$, then the phase can be
approximated as $\omega t- kz\approx \omega t$. The consequence is
that the magnetic field is neglected, since magnetic field is
given by $\textbf{B}=\nabla \times \textbf{A}$. If the laser field
is sufficiently intense, then a fully relativistic treatment for
the dynamics of the electrons is
required. Since the wave length is not small, the dipole approximation  fails for the intense laser fields. \\
As it was discussed in \cite{enson}, there is, however, an
intensity region intermediate between the nonrelativistic and
relativistic domain. The force on a charged particle due to the
magnetic field is related to $\ds{\textbf{v}/c}$. This can be seen
from the Lorentz force formula
$\ds{\textbf{F}=e(\textbf{E}+\textbf{v}/c\times \textbf{B})}$.
True relativistic effects, however, are $\ds{v^2/c^2}$ effects.
Hence, there is an intermediate region where the dipole
approximation is no longer valid and the relativistic treatment is
not necessary. In this region, the particle is described by
Schrodinger equation which includes all relativistic effects of
order $v/c$. Some methods have been developed to understand the
dynamics of laser-matter interaction in this region
\cite{enson,enson2,enson3,enson4,enson5}. The exact solution
obtained here can also be used for the domain where $v/c$ but not
$v^2/c^2$ are necessary.\\
Finally, we think that the analysis of high harmonic generation
\cite{HHG,HHG2} can be performed more accurately by using the exact solution (\ref{cozum20}).\\

\end{document}